\documentclass{article}
\usepackage{setspace}
\usepackage[margin=1.25in]{geometry}
 
\usepackage{amsmath,amsthm,amssymb}
\usepackage{verbatim}
\usepackage{relsize}
\usepackage{listings}
\usepackage{xcolor}
\usepackage{hyperref}
\usepackage{graphicx}

\title{Subgroup Difference in Differences to Identify Effect Modification Without a Control Group}
\author{Zach Shahn}
\date{June 2023}

\begin{document}

\maketitle

This note presents a very simple identification result that I highly doubt is original. It is so straightforward that somebody has likely thought of it previously. However, possibly due to poor search strategies, I have not been able to find the result in the literature. So I will present it here, and if somebody points me to where it was previously published I will take this note down.\\

Suppose you wish to assess the extent to which a new policy had different impacts in different subgroups of interest (as defined by baseline covariates) using only pre-post data from a sample of those impacted by the policy, i.e. with no control group. For example, suppose only one state in the country collects data on a given outcome, e.g. some measure of quality of life. That state then increases its minimum wage, and you wish to assess whether the wage increase had different impacts on the quality of life measure in men and women. We will show that under the assumption that the counterfactual untreated outcomes in men and women followed parallel trends pre- and post- intervention (`subgroup parallel trends'), a simple difference in differences (DiD) expression subtracting the average pre-post outcome difference in men by the average pre-post outcome difference in women identifies the difference between the policy's effects on men and women.\\ 

This identification result for effect modification by a baseline covariate (sex in the example above) holds even if we cannot identify the conditional treatment effect at any level of the baseline covariate. I can imagine several scenarios in which it might be of interest to estimate the magnitude of effect modification by baseline covariates even in the absence of identification of conditional causal effects. One setting is when there are concerns surrounding equity, and investigators might wish to ensure that sensitive subgroups are not benefiting less from an intervention. In the minimum wage example above, one might want to ensure that the policy did not increase the wage gap. Another setting is if a substantive theory or model makes predictions about the relative impacts of a policy on different subgroups, and investigators wish to test the theory. In our running example, perhaps some social theory predicts that one sex would have their quality of life impacted to a greater extent on average by increased wages, and the goal of the analysis is to test that theory.\\

Given availability of a control group, Abadie (2005) describes how to estimate heterogeneous effects under a parallel trends assumption across groups. 

To formalize the problem, assume we observe independent identically distributed realizations of the random variable $O=(X,Y_{0},Y_{1})$, where $X$ denotes a baseline covariate, $Y_0$ denotes pre-treatment outcome at time $0$, and $Y_1$ denotes post-treatment outcome at time $1$. All units are treated at time $1$. Let $Y_1(0)$ denote the counterfactual untreated outcome value (Rubin, 1974) at time $1$. Let $Y_1(1)$ denote the potential outcome under treatment at time $1$. We make the standard consistency assumption that 
\begin{equation}\label{consistency}
Y_1(1) = Y_1,
\end{equation}
as all subjects receive treatment at time $1.$ Interest centers on effect modification by $X$, i.e. 
\begin{equation}\label{estimand}
E[Y_1(1)-Y_1(0)|X=x] - E[Y_1(1)-Y_1(0)|X=x']
\end{equation}
for values $x$ and $x'$ of $X$. We make the subgroup parallel trends assumption across levels of $X$, i.e.
\begin{equation}\label{parallel_trends}
E[Y_1(0)-Y_0|X=x] = E[Y_1(0)-Y_0|X=x'],
\end{equation}
for values $x$ and $x'$ of interest. We can then easily show that the subgroup difference in differences (SDiD) expression
\begin{equation}\label{did}
E[Y_1-Y_0|X=x] - E[Y_1-Y_0|X=x']
\end{equation}
identifies the causal estimand (\ref{estimand}) under subgroup parallel trends assumption (\ref{parallel_trends}). The proof is as follows:
\begin{align}
\begin{split}\label{derivation}
    &E[Y_1-Y_0|X=x] - E[Y_1-Y_0|X=x']\\
&=(E[Y_1-Y_0|X=x]-E[Y_1(0)-Y_0|X=x]) - (E[Y_1-Y_0|X=x']-E[Y_1(0)-Y_0|X=x'])\\
&=E[Y_1-Y_1(0)|X=x] - E[Y_1-Y_1(0)|X=x']\\
&=E[Y_1(1)-Y_1(0)|X=x] - E[Y_1(1)-Y_1(0)|X=x']
\end{split}
\end{align}
(\ref{parallel_trends}) implies that the second line of the derivation is obtained from the first line by adding and subtracting the same quantity. The third line follows from canceling out $Y_0$ in the second line. The last line (which is equal to the causal estimand (\ref{estimand})) is equal to the third line by consistency (\ref{consistency}).\\
 
 Plug-in estimation of (\ref{estimand}) based on (\ref{derivation}) is straightforward. If $X$ is categorical, empirical sample averages would probably suffice. Otherwise, if $X$ is continuous, estimation would require a regression model for $E[Y_1-Y_0|X]$. Note in the case of continuous or high dimensional categorical $X$ that separate subgroup parallel trends assumptions are required for any two levels of $X$ to be compared. \\

Plug-in estimation based on (\ref{derivation}) is just the difference of pre-post comparisons within the subgroups of interest. It is exactly what a naive analyst would do if they believed simple pre-post comparisons would yield valid effect estimates and they were also interested in heterogeneous effects. For this reason, I am sure that such estimators have been computed countless times. However, I am not sure if it has been previously noted that this obvious approach is valid under the subgroup parallel trends assumption (\ref{parallel_trends}) even if the pre-post comparisons do not yield valid causal effects for each subgroup.\\

Of course, the subgroup parallel trends assumption (\ref{parallel_trends}) is extremely strong and untestable. One of its implications is that the covariate of interest $X$ is not a time-modified confounder in the sense of Platt et al (2009). However, it is not stronger (or weaker) than the standard DiD assumption of parallel trends across treatment groups as opposed to covariate values. Abadie (2005) describes how to characterize effect heterogeneity given a baseline covariate if a control group is available and the standard parallel trends assumption across treatment groups holds. Many analysts would happily conduct a traditional DiD study along the lines of Abadie (2005) but throw up their hands in the absence of a control group due to the typically less plausible assumptions required by interrupted time series designs (e.g. Chapter 6 of Cook et al, 2002). Given a successful pre-trends test across levels of the baseline covariate $X$, such analysts might wish to explore effect modification via SDiD instead of giving up altogether. This is particularly the case if effect modification is of primary interest, as in the equity or theory-testing scenarios mentioned above.\\

Finally, we note that SDiD estimates of effect heterogeneity may be preferred to traditional DiD estimates of effect heterogeneity (e.g. using the estimators in (Abadie, 2005)) even if a control group is available. This is because the subgroup parallel trends assumption within the treated group may sometimes hold when the standard parallel trends assumption does not. In particular, suppose other circumstances change in the treated group (but not the control group) at the same time the intervention is implemented. Then standard parallel trends would not be expected to hold. But if those additional changes (perhaps in contrast to the intervention of interest) are expected to impact the two subgroups of interest similarly, then the subgroup parallel trends assumption may still hold.
\section*{Bibliography}
Abadie, Alberto. "Semiparametric difference-in-differences estimators." The review of economic studies 72, no. 1 (2005): 1-19.\\
\\
Cook, Thomas D., Donald Thomas Campbell, and William Shadish. Experimental and quasi-experimental designs for generalized causal inference. Boston, MA: Houghton Mifflin, 2002.\\
\\
Platt, Robert W., Enrique F. Schisterman, and Stephen R. Cole. "Time-modified confounding." American journal of epidemiology 170, no. 6 (2009): 687-694.\\
\\
Rubin, Donald B. "Estimating causal effects of treatments in randomized and nonrandomized studies." Journal of educational Psychology 66, no. 5 (1974): 688.

\end{document}